\documentclass[10pt]{article}

\pdfoutput=1

\usepackage{amsmath}
\usepackage{amssymb}

\usepackage{graphicx}
\usepackage{tipa}
\usepackage{times}
\usepackage{hyperref}

\usepackage{cite}

\usepackage{color} 

\usepackage[labelfont=bf,labelsep=period,justification=raggedright]{caption}

\usepackage{soul}
\usepackage{array}
\newcolumntype{x}[1]{|>{\centering\arraybackslash\hspace{0pt}}p{#1}|}

\makeatletter
\renewcommand{\@biblabel}[1]{\quad#1.}
\makeatother

\date{}

\begin{document}

\begin{flushleft}
{\Large
\textbf{Internal and external dynamics in language: Evidence from verb regularity in a historical corpus of English}
}
\\
*Christine F. Cuskley$^{1}$, 
Martina Pugliese$^{2,1}$, 
Claudio Castellano$^{1,2}$,
Francesca Colaiori$^{1,2}$,
Vittorio Loreto$^{2,3,1}$,
Francesca Tria$^{3}$
\\
\bf{1} Institute for Complex Systems (ISC-CNR), Via dei Taurini 19, 00185 Roma, Italy
\\
\bf{2} Sapienza University of Rome, Physics Dept., P.le Aldo Moro 5, 00185 Roma, Italy
\\
\bf{3} Institute for Scientific Interchange (ISI), Via Alassio 11C, 10126 Torino, Italy
\\
$\ast$ E-mail: ccuskley@gmail.com
\end{flushleft}

\section*{Abstract}
Human languages are rule governed, but almost invariably these rules have exceptions in the form of irregularities. Since rules in language are efficient and productive, the persistence of irregularity is an anomaly. How does irregularity linger in the face of internal (endogenous) and external (exogenous) pressures to conform to a rule? Here we address this problem by taking a detailed look at simple past tense verbs in the Corpus of Historical American English. The data show that the language is open, with many new verbs entering. At the same time, existing verbs might tend to regularize or irregularize as a consequence of internal dynamics, but overall, the amount of irregularity sustained by the language stays roughly constant over time. Despite continuous vocabulary growth, and presumably, an attendant increase in expressive power, there is no corresponding growth in irregularity. We analyze the set of irregulars, showing they may adhere to a set of minority rules, allowing for increased stability of irregularity over time. These findings contribute to the debate on how language systems become rule governed, and how and why they sustain exceptions to rules, providing insight into the interplay between the emergence and maintenance of rules and exceptions in language. 

\section*{Introduction}
Language is a continuously evolving system, subject to constant pressures to enhance expressivity while guaranteeing successful communication over time. William D. Whitney~\cite{Whitney_1875}, one of the earliest American English lexicographers, highlighted that language can be seen as a open system subject both to {\em conservative} and {\em alterative} forces. In this view, language is a living system continuously experiencing birth, growth, decay, and death. Accordingly, language changes as a result of external or exogenous pressures, but is also subject to internal or endogenous pressures.

Endogenous factors are related to inner instabilities of languages leading to changes independent of external disturbances: for instance, in phonology, pressure against ambiguity makes sounds maximally contrastive to decrease the likelihood of misunderstanding~\cite{Liljencrants_Lindblom_1972,Lindblom_1986}. More generally, any mistakes or unconscious innovations introduced by speakers could trigger endogenous change in language. On the other hand, exogenous factors are related to historical, political, social, and technological factors that affect language communities~\cite{Labov_1972,Muhlhausler_1996,Mufwene_2001}. For instance, an exogenous influx of non-native speakers into a language has been correlated with decreased morphological complexity~\cite{lupydale}.

Despite a huge literature concerning language change and its causes (e.g., ~\cite{McMahon1994,Croft2000}), change is often considered in terms of \textit{either} exogenous \textit{or} endogenous pressures (although see ~\cite{JonesEsch2002}), and quantitative accounts of their combined effects have only recently begun to emerge ~\cite{GriesHilpert2010,Hilpert2013}. The recent availability of large-scale historical corpora, and the use of a more quantitative complex systems approach, has allowed ever increasing detail in documenting language change, from the trajectories of individual words~\cite{Michel_2011,Perc2012} to the broad statistical properties of language over time~\cite{Petersen_2012a,Petersen_2012}.

Here we leverage this opportunity by examining how endogenous and exogenous pressures affect the interplay between regularity and irregularity in language. To this end, we study the formation of the simple past tense in American English, contrasting regular forms (e.g., \textit{look-looked}) with irregular constructions (e.g., \textit{ring-rang}). The past tense has been considered extensively as a general exemplar for how language is cognitively structured~\cite{Pinker_Ullman_2002}, and also as an illustration of how high frequency linguistic variants exhibit greater stability ~\cite{lieb2007,carrol2012,Bybee2007}. Here we take a diachronic perspective, monitoring how regular and irregular forms emerge, stabilize, and change over time. The Corpus of Historical American English (CoHA~\cite{davies2012}) is composed with a balanced variety of written genres, including more than 330 million words from the 1830-1989 period, each tagged for part of speech.

CoHA allows us to define regularity in terms of actual usage, giving a proportion of irregular usage for each verb at any given time.  We observe variations in regularity of clear exogenous or endogenous nature, coming both from birth (exogenous) and regularization/irregularization (endogenous) processes.  We highlight how exogenous and endogenous processes have different impacts on the dynamics of regularity. The expansion of the vocabulary drives the increase in regularity through the entrance of new, mostly regular, verbs. However, at odds with previous findings from historical corpora ~\cite{lieb2007}, the endogenous process of regularization does not foster significant changes in overall regularity. Endogenous regularization processes, in fact, turn out to be paralleled by comparable irregularization processes, leaving the number of irregular verbs roughly constant over time.  

All this implies that the relationship between verb instability and frequency is more complex than previous corpus studies have indicated ~\cite{lieb2007}. In particular, we show that sound similarity plays a crucial role in the persistence of irregularity. By defining classes of irregular verbs based on sound similarity, the relationship between irregularity and frequency is made clearer. We show that the trajectories of individual verbs and classes over time allow insights into the underlying rule set, informing the longstanding debate regarding whether language is strictly rule-governed with a static set of exceptions~\cite{Pinker_1999}, or a set of connectionist processes leading to a system that appears to have explicit rules~\cite{rumel1986,mclelpatterson2002}. Rather than either of these extremes, the past tense is comprised of  competing set of rules, with the regular rule dominating the majority of verb types, and a collection of minority rules applying to irregular verbs~\cite{Yang_2002}. Our findings show that regularities and exceptions emerge from the dynamic interplay of exogenous and endogenous forces: a push and pull between the external influx of new verbs and a struggle among existing verbs to hit the moving target of a dynamic rule set.

\section*{Results}

Our analysis aims at presenting a quantitative and balanced picture of factors contributing to language change. Focusing on verbs as our testing ground, we document the effects of verb birth and death, as well as the transformation of existing words (regularization/irregularization). In order to analyze the effect of the different factors we separate our analysis in two parts. First, we focus on the full dataset spanning 16 decades (1830-1989) in order to characterize the temporal dynamics of an open language system. In this case, the ensemble of verbs form an \textit{extended vocabulary} whose size and composition vary due to verbs entering and exiting the language (data available in file S7). Secondly, we focus on changes occurring in the \textit{core vocabulary} considering only verbs which occur in each of the 16 decades considered.  We refer the reader to the Materials and Methods section and the Supporting Information for details on corpus preparation (S3 in file S1) and data analysis (S4 in file S1).

\subsection*{Language as an open system}

\begin{figure}[h!]
  \includegraphics[width=\textwidth]{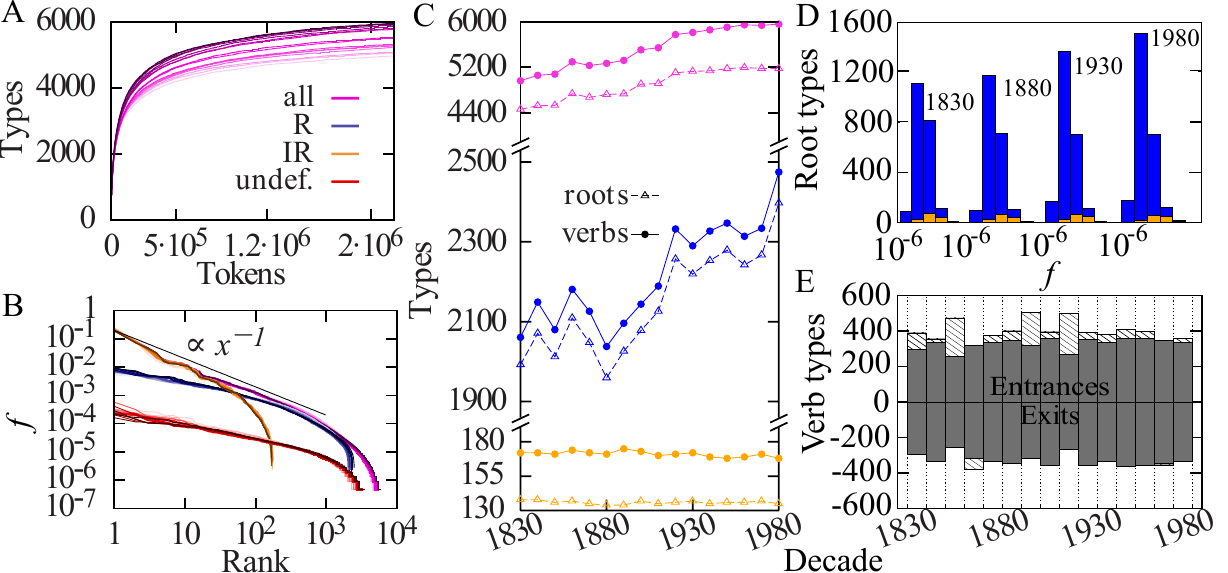}
  \caption{{\bf Language as an open system.} Colors indicate the type of verb or root depicted: all verbs (purple), mostly regular (blue), mostly irregular (orange), and undefined (red). Lighter saturation of a colour indicates earlier decades, while darker saturation indicates later decades. Where relevant, decades are labelled with the first year, e.g., 1830-1839 is 1830. (A) The number of types versus the number of tokens for all verbs in each decade. Given an equal number of tokens, the increase in the number of types over time reflects a genuine increase in the number of verbs at fixed sample size. (B) Frequency-rank plot contrasting different kinds of verbs (all, mostly regular, mostly irregular, and undefined). Mostly irregular types constitute the highest frequency regime, undefined types constitute the lowest frequency regime. (C) Plot of the number of mostly regular types, mostly irregular types, and all types over time, both for verbs and roots. The number of mostly regular verb types or root types increases over time, while the number of mostly irregular verb and root types remains essentially constant. The sum of all types considered together is more than the sum of mostly regular and irregular types; this is due to this category encompassing undefined types. (D) Frequency distribution of mostly regular, mostly irregular, and undefined categories for root types in four decades, the starting point of the first frequency bin is indicated for each depicted decade. (E) Verb type birth and death by decade. Each bar refers to the entering (top) and exiting (bottom) number of verb types between two decades; lighter areas show the entrance/exit differential. E.g., the first bar depicts the exits after 1830-1839 and the entrances in 1840-1849. There are generally more verbs entering than exiting, indicative of the overall growth in the number of types.}\label{fig:1}
\end{figure}

The extent to which language can be seen as an open system can be quantified by looking at the growth in the number of distinct verbs (\textit{types}) as a function of the total occurrences of verbs (\textit{tokens}), the equivalent of Heaps' law~\cite{heaps_1978}. Here, in order to highlight the change in language composition over time, we focus on the different decades separately, considering the same number of verb tokens in each decade  (in order to consider genuine growth, rather than increased availability of material for digitized corpora over time; see S3 and Figure S2 in file S1 for further detail). In each decade, types show a sublinear growth as a function of tokens (Fig.~\ref{fig:1}A), implying that the rate at which new verb types enter the vocabulary decreases with the corpus size. The Heaps' curves for different decades do not overlap; rather, they show a clear increase in the number of verb types in consecutive decades, even with a fixed number of tokens (for full curves, see Figure S1 in file S1; for removal information see Table S1 in file S1). This growth indicates a genuine influx of new verbs entering the language in the time period. Figure S1 (file S1) shows the number of tokens prior and after confinement in each decade.

In order to examine changes in the verb system it is convenient to give an operational definition of regular and irregular verbs.  To this end, for each verb and in each decade, we consider the the proportion of irregularity ($I$) as the number of irregular past tense (non {\em -ed}) tokens divided by the total number of past tense tokens.  For each decade, we count types which are mostly regular ($R$) (or mostly irregular, $IR$) as those which have $I<0.5$ ($I \geq 0.5)$.  We also provide separate counts for verb types and root types, the latter combining verbs with the same root to level the effects of individual verb productivity (e.g., {\em do} and {\em undo} are separate verb types, but instances of a single root type; in file S1, see S2, Table S2, Table S3). Many types do not appear with sufficient frequency to be definitively classified as regular or irregular; such types are categorized as \textit{undefined} (i.e., appear with zero or extremely low frequency of usage in the past tense; in file S1, see S2, Table S2, Table S3). 
Figure S3 (in file S1) shows, for each category of root types, the histogram in frequency in four different decades (data for root types is available in file S8).
Fig.~\ref{fig:1}B complements the Heaps-like plot by showing frequency-rank plots~\cite{zipf_1949} for different verb types in all decades, showing that the high frequency regime is dominated by irregulars.  We now turn to the dynamic properties of the corpus, by explicitly showing how the total number of types grows over time.  Fig.~\ref{fig:1}C shows the number of verb types and root types across decades, for mostly regulars, mostly irregulars, and all types. While there is no significant variation in the number of mostly irregular types, mostly regular types display a clear increasing trend, indicating that entering types are predominantly regular.

The activity and growth of the vocabulary is even more evident in Fig.~\ref{fig:1}E, where the number of verb types entering or exiting the language in each decade is plotted. The imbalance between entrances and exits (lighter areas) gives rise to the overall growth of the number of verbs over time.  Furthermore, Fig.~\ref{fig:1}D shows the frequency distribution of different kinds of root types, indicating that the growth in the number of types is most prominent for regulars. We can thus conclude that the increase in regularity is driven by an exogenous process of regular verbs entering the language, making for an increase in the ratio of regular types to all verb types as the language evolves.

\subsection*{Dynamics of the core vocabulary}

We now turn our attention toward the endogenous processes of regularization and irregularization. To this aim, we restrict ourselves to types which occur at least once in each of the 16 decades.  In addition, in order to capture the genuine dynamics of regularization/irregularization (rather than word formation processes), in the core vocabulary we consider exclusively root types. We thus define the core vocabulary as the set of all (3596) root types which occur at least once in each of the 16 decades. For the present analysis, root types are defined as regular (irregular) in a given time period if their proportion of irregularity $I$ is equal to $0$ ($1$) within a tolerance threshold, $\epsilon$ (see S4, file S1). Thus we can further identify {\it stable regulars} ({\it stable irregulars}) if they are regular (irregular) in all decades. 
All other core roots, which have $0<I<1$ in at least one decade, are considered {\it active}. This latter set includes roots for which some regularization or irregularization process is in progress, or roots whose regular past tense form coexists with an irregular form. Note that although core roots are by definition present in each decade with some overall frequency, some root types have zero or few occurrences in the past tense, so that their regularity can be undefined in some decades (see file S1: S4, Table S2 and Table S3 for further detail).

The dynamics observed in the core vocabulary are the result of endogenous regularization/irregularization processes only. The first observation is that the vast majority of types are stable regulars or irregulars (types which are regular or irregular in every decade, respectively), and only exhibit fluctuations in frequency ($f$). In the midst of these two classes of stable root types, there is a minority of 81 active roots, whose $0<I<1$ in at least one of the sixteen decades considered.

\begin{figure}[h!]
\begin{center}
\includegraphics[width=10cm]{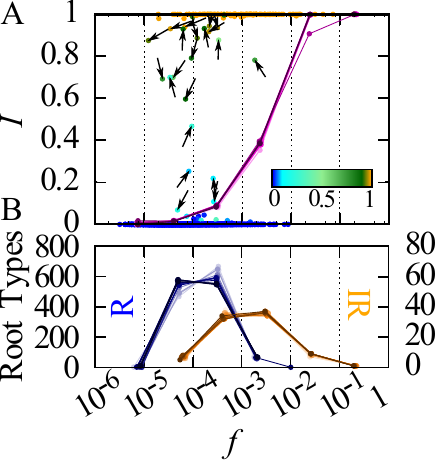}
\end{center}
  \caption{{\bf Relationship between frequency $f$ and regularity in the core vocabulary.} Lighter saturation of a color indicates earlier decades, while darker saturation indicates later decades. (A) Single points represent roots in the final decade (1980-1989), color-coded according to the average value of $I$ across all 16 decades. Arrows are placed on 21 active verbs with $0.05 < I < 0.95$. The direction of each arrow indicates the trajectory from the first to the last defined occurrence in the ($f,I$) space; the length of arrows is fixed. The purple curves show the binning of both $f$ and $I$ in each frequency logarithmic bin, for each decade. (B) Distributions of the number of stable regular roots (left axis) and stable irregular roots (right axis) for each decade, showing separate peaks which indicate different frequency profiles.}\label{fig:2}
\end{figure}

Fig.~\ref{fig:2}A shows $I$ as a function of $f$ for each of the core roots, considering these values in the last decade (1980-1989) plus the binned values in each decade, which demonstrate that the observed behavior is qualitatively similar across all the decades considered.  As already stressed, the overwhelming majority of verbs have $I=0$ or $I=1$, and as other studies have pointed out~\cite{bybee1985,lieb2007}, irregular roots are generally higher in frequency than regular roots. Almost all active root types span a frequency interval between $10^{-5}$ and $10^{-2}$. Above this range mainly high frequency irregular roots are present, while below this range predominantly regular roots are observed.  The arrows on the active roots indicate the trajectory of the root in the ($f,I$) plane from the first decade where the root was defined (i.e., an arrow pointing up indicates an increase in $I$ since the first decade, an arrow pointing right indicates an increase in $f$ since the first decade, and so on). The observed patterns remain remarkably constant over time, dominated by the stable regulars and irregulars. As expected, the binned value of $I$ turns out to be a growing function of frequency $f$, passing from approximately 0 for $f<10^{-4}$ to approximately 1 for $f>10^{-2}$.  However, the transition from regulars to irregulars is shifted towards high frequencies with respect to the frequency band where we mainly observe active types. This is due to the larger overall number of regular types dominating the irregularity proportion in each bin. Fig.~\ref{fig:2}B shows that the frequency distributions of stable regular and stable irregular root types are well separated.

\begin{figure}[!ht]
\includegraphics[width=\textwidth]{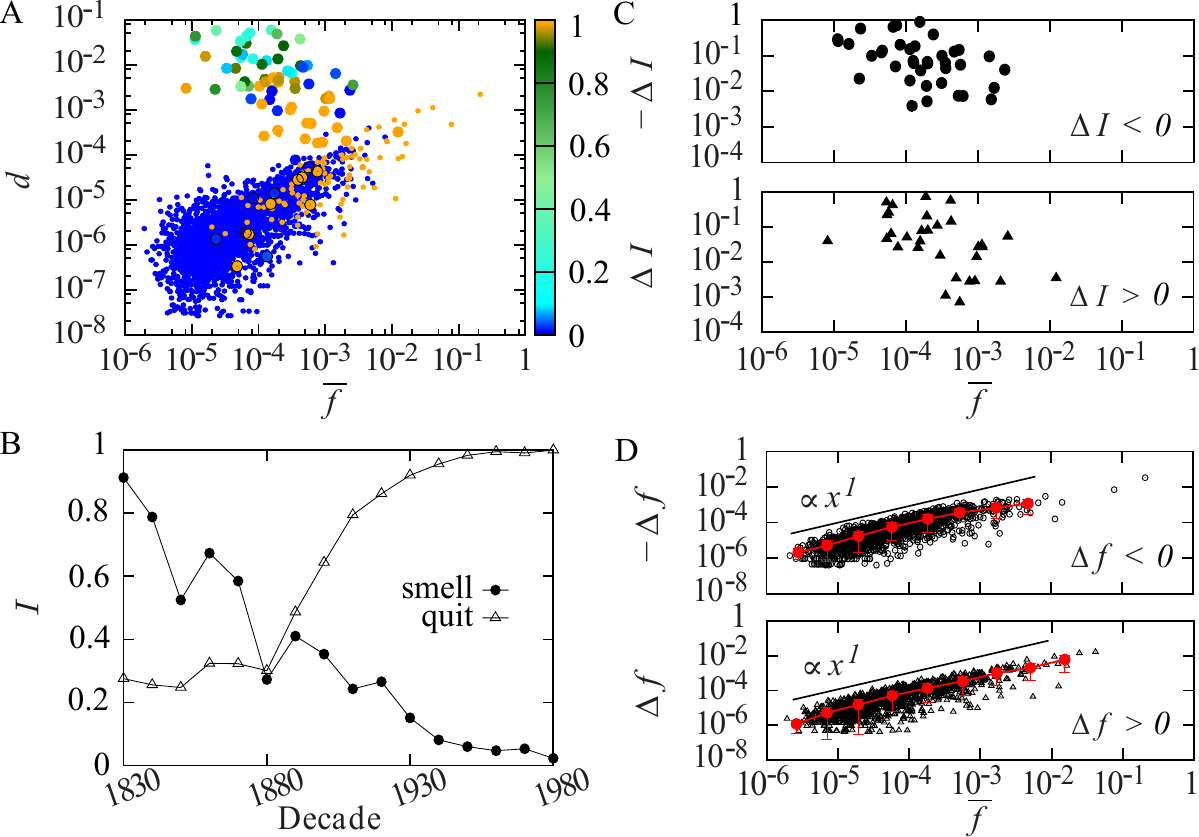}
\caption{
{\bf Excursions in frequency and proportion of irregularity in the core vocabulary.} 
  (A) Plot of $d$ against $\bar f$ (frequency averaged across all the decades) for core roots. Points are color-coded according to their average $I$ across all the decades. Active roots are represented by larger points, patterning predominantly in the upper ``cloud,'' distinctly away from stable regular and irregular roots.
Some active roots fall into the lower ``cloud'' (indicated by black circles), due to an effective net $\Delta I = 0$ (i.e., $I$ fluctuates below 1 or above 0 between the first and last decade, but has the same value in the first and last decade). 
(B) Plot of the time evolution of $I$ for exemplars of irregularization (\textit{quit}) and regularization (\textit{smell}).  (C) Excursion in irregularity proportion $\Delta I$ ($-\Delta I$ when negative) calculated as the difference in $I$ between the first and last decade plotted against $\bar f$ for the 69 active roots with a non-zero $\Delta I$ between the first and last occurrences. A negative $\Delta I$ indicates regularization, a positive $\Delta I$ indicates irregularization. The decreasing trend with increasing average frequency corresponds to the behaviour of the ``cloud'' in (A).  (D) Excursion in frequency $\Delta f$ 
($-\Delta f$ when negative) against $\bar f$ for all core roots. Red points represents binning of $\bar f$ and $\Delta f$, error bars are given for $\Delta f$. The trend of $|\Delta f|$ proportional to $\bar f$ is visible.}
\label{fig:3}
\end{figure}

The dynamics of roots in the core vocabulary are more deeply understood by measuring the excursion of each root type in the $(f,I)$ plane, defined as $d=\sqrt{(\delta I)^2+(\delta f)^2}$, where $\delta I$ and $\delta f$ represent the variations of $I$ and $f$ divided by the number of decades where the root is defined ($\Delta I$ and $\Delta f$ refer to the total change). Fig.~\ref{fig:3}A shows $d$ as a function of $\bar f$, the average frequency across all the decades, for all core root types; points are coloured according to the average $I$ over time.  Interestingly, points are clustered in two different regions, a signature of the emergence of two distinct behaviors. Active types are concentrated in the upper cloud, characterized by a decreasing trend of $d$ in frequency. Here the variation of $d$ is mostly due to variation of $I$ (see also Fig.~\ref{fig:3}C), although 12 of the active verbs have a $\Delta I$ of $0$ between the first and last decade (though by definition, they have some fluctuation in the intervening decades).  On the other hand, stable regulars and irregulars have $d = |\delta f|$ roughly proportional to $f$ (see also Fig.~\ref{fig:3}D).  In order to disentangle the contributions of frequency and irregularity to $d$, in Fig.~\ref{fig:3}C and~\ref{fig:3}D we plot $\Delta I$ and $\Delta f$ separately as a function of $\bar f$; $\Delta I$ for active roots only (\ref{fig:3}C) and $\Delta f$ for all core roots (\ref{fig:3}D).

Interestingly, both the $\Delta I$ and $\Delta f$ curves appear roughly symmetric with respect to positive and negative variations. As for irregularity variations, the number of verbs changing significantly their $I$ value, and the amplitude of these variations, are roughly the same for regularization and irregularization, indicating 
a balance in the two processes. Moreover, the changes in $I$ appear to be similar in both directions (towards regularization and towards irregularization) for any fixed value of $\bar f$, again pointing to the presence of opposite endogenous forces of comparable strength. 
A similar situation is observed for frequency variations. In particular, while the amplitude of frequency changes seems to be strongly correlated with frequency itself, the direction of the variation appears independent from it, and roughly the same number of verbs increase and decrease their frequency at any given value of $\bar f$. Overall, six roots exhibit dramatic changes in $I$ over the 16 decades, with three verbs fully regularizing and three fully irregularizing: \textit{smelled, spilled} and \textit{spelled} overtake \textit{smelt, spilt} and \textit{spelt}, while \textit{quitted, lighted} and \textit{wedded} are replaced by \textit{quit, lit} and \textit{wed}. Fig.~\ref{fig:3}B shows the proportion of irregular usage $I$ of \textit{quit} and \textit{smell} over time, showing the complete irregularization of \textit{quit} and the regularization of \textit{smell} (see Figure S5, file S1 for the other verbs).

\subsection*{Relevance of exogenous and endogenous factors}

Contrasting the extended vocabulary with the core vocabulary highlights two different types of dynamics in the language.  The considerable expansion of the extended vocabulary -- new verbs entering the language -- is clearly an exogenous process.  The dynamics of root types in the core vocabulary exemplify language internal, endogenous processes.  Both processes have the potential to affect regularity; however, exogenous dynamics are far more influential in this regard.

The evidence presented so far clearly indicates that overall, the set of verbs tends to become more regular over time: the ratio of mostly irregular root types to mostly regular root types ($y$) has decreased from $y_{in} \approx 7.1\% (142/1,992)$ in 1830-1839 to $y_{fin} \approx 5.7 \% (137/2,397)$ in 1980-1989.  This is due to the considerable growth in the number of regulars, while the number of irregulars remains essentially constant throughout 160 years.  The contribution of endogenous regularization of existing verbs to the growth in regularity is minimal: the driver of the global regularization process is instead the exogenous birth of new regular types.

The exogenous effect of new regular verbs entering the language is quantitatively overwhelming with respect to that of endogenous regularization.  However, while regularity increases in terms of types, it remains constant at the system level. In other words, the total proportion of irregular usage $I_{tot}$ is fairly constant over time: between $65-70\%$ of all past tense verb tokens are irregular (see Figure S6, file S1).  Although many new verbs enter the system as regulars, they do so at very low frequency, so that their impact on $I_{tot}$ is exceedingly small.  High-frequency types, which constitute the bulk of verb usage, tend to be irregular and remain so over time.

\subsection*{Drivers of regularization and irregularization}

Although our data show that the core of the vocabulary is stable overall, there remains a minority of active, dynamic verbs within the language, undergoing processes of regularization or irregularization. Here, we consider what causes this activity from an endogenous perspective.

A good deal of consideration has been given to why regularization occurs. From a system perspective, there may be pressure for individual verbs to conform to the dominant rule~\cite{Pinker_Ullman_2002}, creating a system with a more concise overall rule-set attractive to learners~\cite{KirbyCornishSmith2008}. From a learner's perspective, regularization occurs as an individual fails to accurately retrieve an irregular form, and falls back on the dominant regular rule~\cite{albrighthayes2003}. The phenomenon of irregularization has been given less consideration, as it is often considered so rare as to be irrelevant~\cite{lieb2007,Pinker_2000}. However, our data show that in terms of types, irregularization is roughly as prevalent a process as regularization; but what cognitive and system pressures might account for it?

Phonological analogy may be a powerful process in this regard~\cite{bybee2001,skousen1989}. Although some of the most frequent irregulars are suppletive (i.e., there is little connection between the present and past form, e.g.,\textit{be/was}), many irregulars form the past by application of systematic changes to their roots, and share striking similarities with other irregulars (e.g., \textit{sing, ring} $\rightarrow$ \textit{sang, rang})~\cite{mclelpatterson2002}. In other words, highly frequent irregular verbs may form attractors, which have the potential to draw phonologically similar regulars. This process could account not only for processes of irregularization, but also for the presence of some anomalous stable irregulars with relatively low frequency. Regularity, on the other hand, is a rule which applies at the morphological level, and thus, regulars do not act as strong attractors for other phonologically similar verbs (however, the regular rule does exhibit phonologically conditioned variation, as in the difference in the spoken realization of \textit{-ed} in \textit{walked} vs. \textit{faded}; ~\cite{albrighthayes2003}). Rather, regularization occurs as a consequence of broad application of a morphological rule, which is applied indiscriminately when an irregular exception is unknown, and this becomes more likely as frequency declines.

To operationalize this perspective, we propose a clustering of the set of irregular root types into classes based on phonological similarity.  We group all roots which have some $I > 0$ into classes based upon the type of change which occurs to the infinitive to form the past tense. For example, \textit{ring} transforms to \textit{rang}, placing it in the same class as \textit{sing} (see Table S4, file S1 for full listing of classes). Each class has the potential to draw new irregular members over time, but may also lose low frequency members to regularization (see S5, file S1).

\begin{figure}[!ht]
\includegraphics[width=\textwidth]{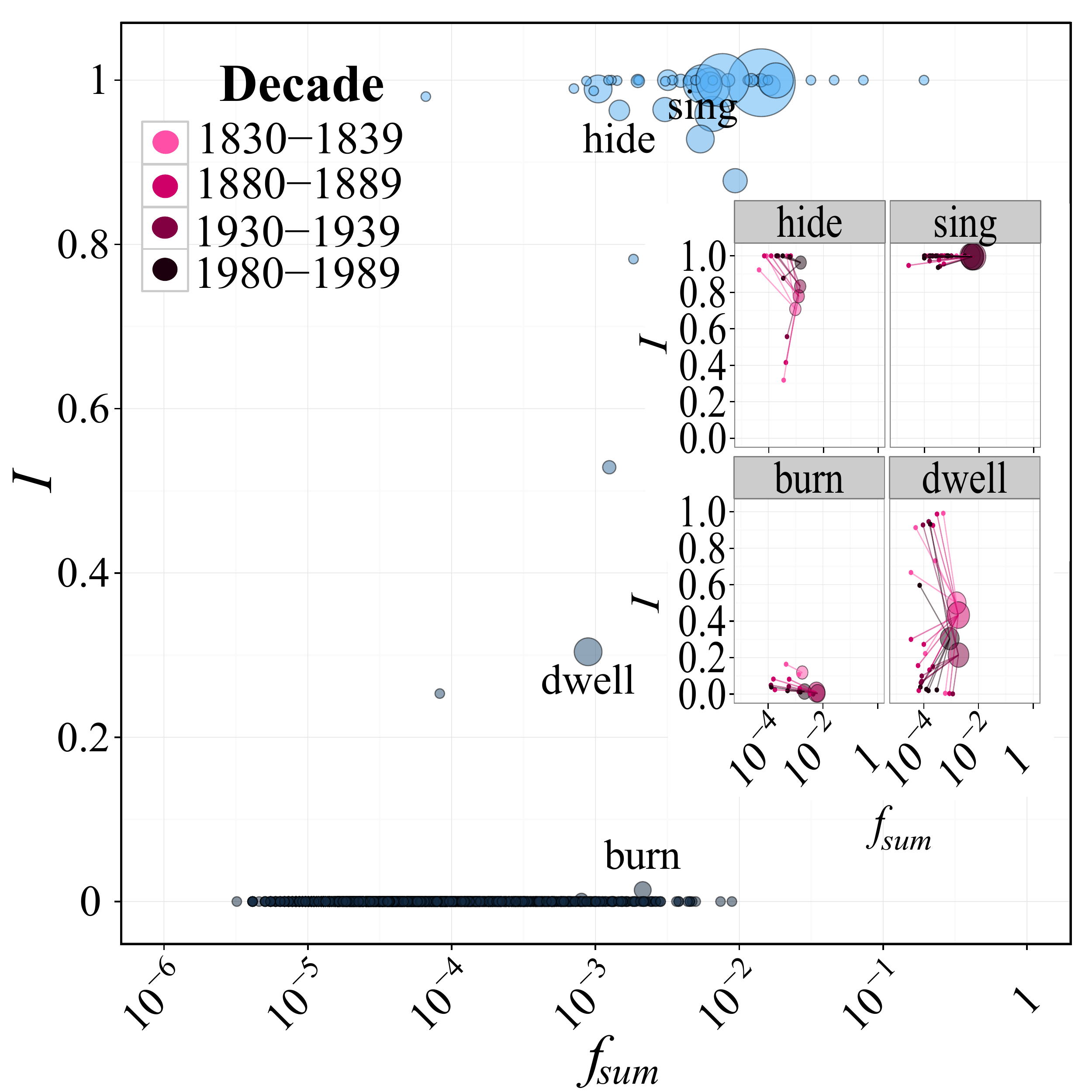}
\caption{
{\bf Plot of the proportion of irregularity, $I$, for phonological classes against  $f_{sum}$, the sumed frequencies of members in a class.} The main plot refers to the last decade (1980-1989) and, along with classes, shows regular roots in the decade (grey points at the bottom). For each unclassed regular root, the same values of $I$ and $f$ are used as in Fig.~\ref{fig:2}A. The size of the circle for each class is proportional to the number of members in the class. The insets show four exemplars of classes with different behaviours, in four time snapshots (identified with a purple hue going from light to dark with increasing time): the largely irregular \textit{hide} and \textit{sing} classes on the top, and the regularizing \textit{burn} and \textit{dwell} classes on the bottom. Small points in the insets are the
member roots of the class (with their values of $I$ and $f$), connected with lines to the class itself. 
The relationship between variance of $I$ within a class and its stability is visible especially in the "star-like" quality of the \textit{dwell} class.
The plot shows that even basic phonological classification makes the frequency/regularity relationship clearer, as compared to Fig.~\ref{fig:2}A.}
\label{fig:4}
\end{figure}

Nothing guarantees that this criterion provides an optimal classification, and the quest for irregular classes which align most closely with the actual grammars of speakers is a challenging open problem. However, even this basic classification has the effect of clarifying the otherwise anomalous behavior of some verbs, and making the separation in frequency between regulars and irregulars more distinct by shifting lower frequency irregulars. Fig.~\ref{fig:4} displays $I$ for a given class (calculated as all non \textit{-ed} tokens in a class over the sum of all past tokens) versus frequency (summed over all members) in the final decade. It demonstrates that even a basic phonological classification makes the dependence of regularity on frequency considerably clearer, as contrasted with Fig.~\ref{fig:2}A.

The insets in Fig.~\ref{fig:4} show four representative classes in four snapshots over time. The \textit{hide} and \textit{dwell} classes represent two different states of class instability, while the \textit{sing} and \textit{burn} classes are essentially stable.  The \textit{hide} class is destabilized as it actively draws a new irregular (in this case, \textit{light}), but approaches a more stable state towards the end of the time period as \textit{light} fully irregularizes.  The \textit{dwell} class, on the other hand, represents a highly destabilized class as its members regularize. Concerning more stable classes, the \textit{sing} class is largely stable in its irregular state, and explains the existence of persistent low frequency irregulars such as \textit{spit}.  The \textit{burn} class, on the other hand, is rapidly losing members to regularization, and has almost completely regularized by the end of the time period. Based on these results, the stability of a class seems related to the variance of $I$ among members of a class (see Figure S6, file S1).

\section*{Discussion}

Language is an open system with many different processes acting simultaneously to characterize its dynamics. In our analysis of American English we put a few of these processes in perspective by dissecting the dynamics in verb birth, death, regularization, and irregularization, and estimating their relative relevance. Our study allows for several novel observations about the diachronic dynamics of verb irregularity.

First we observe that, despite the growth of the number of types, the total number of irregular types stays largely constant over time, with a considerable and consistent percentage of tokens (between 65-70\%; see Figure S5, file S1) being irregular past tense forms, due to the irregularity and stability of the most frequent verbs.  This implies that, despite the increase in expressive power, the cognitive effort required to master the structure of past tense forms is constant over time.  

The second important observation comes from the time evolution of the verb system.  Important changes occurred during the 160 years considered, with a substantial growth in the percentage of regular verbs. The origin of this variation is exogenous: a large number of new regular verbs are introduced into the language, many more than those (also typically regular) which get extinct.  Apart from the large influx of verbs entering the language the vast majority of verbs present over all 16 decades remains in the categories of stable regulars or stable irregulars, with stationary distributions well separated in frequency. However, even in this relatively short time period, we observe some endogenous dynamics in the core vocabulary.  Endogenous irregularization, often considered so rare as to be irrelevant, was about as relevant as regularization: the number of regular verbs shifting toward irregular is comparable to the number of irregulars becoming regular. Endogenous dynamics take place within a specific frequency window where transitions in both directions occur, a scenario sharpened by classifying irregular verbs in terms of their phonological similarity.

In conclusion, a diachronic analysis of past tense formation in American English reveals a complex interplay of exogenous and endogenous factors affecting how rules and exceptions coevolve. This points to a more general scenario in which rules in language, far from being fixed over time, emerge, evolve, and compete in a self-organizing way. This self-organization is a response to exogenous pressures exerted by the porous character of the system, as well as the endogenous need to limit cognitive effort while mastering expressivity and comprehension. This work provides a starting point for further studies to investigate the emergence of rules and exceptions in other areas of language, as well as in all the areas of human cognition where norms play a crucial role.

\section*{Materials and Methods}
\subsection*{Corpus Preparation}

CoHA provides part-of-speech tagged frequency lists covering the period between 1810 and 2010 (over 400 million tokens) using the CLAWS set~\cite{garside1987}; we used only the 1830-1989 period (see S3, file S1 for further detail). We began by confining the list from this period only to verbs and lemmatizing it, removing many tagging errors in the process. To avoid potential effects of increasing sample size, we considered an equal number of verb tokens in each decade, according to the size of the first decade considered (1830-1839; 2,177,654 verb tokens). S3 (file S1) provides further detail regarding corpus preparation.

\subsection*{Proportion of Irregularity}

The proportion of irregular usage $I$ for each verb in each decade was calculated by dividing the number of irregular past tense (non \textit{-ed}) tokens in a decade by the total number of past tense tokens for that verb in the decade. Irregular spellings which do not correspond to irregular pronunciation (e.g., \textit{paid} for \textit{pay}) were considered regular tokens. In some cases, a verb or root could have an undefined $I$ for a given decade, due to extremely low frequency (or non-existent) usage in the past tense. See S4 (file S1) for further details.

\subsection*{Phonological Classes}

A full list of all 52 classes and their members is provided in Table S4 (file S1), details about how verb roots were classified are in S5 (both in file S1). Verb roots which exhibited multiple irregular forms were phonologically classified based on their most frequent irregular form (i.e., \textit{swing} was classed with \textit{ring} instead of \textit{string}, although the form \textit{swung} did occur). Suppletive forms such as \textit{do} and \textit{be} were put in their own classes, and forms which did not class identically with any other verbs (e.g. \textit{slay}) were also classed alone.

\section*{Acknowledgments}
This work was supported by the European Science Foundation as part of the DRUST project, a EUROCORES EuroUnderstanding programme.


\clearpage
\section*{Figures}

\clearpage

\section*{Supporting Information}
\begin{itemize}
\item \textbf{S2:} Glossary of terms used for data and analysis.
\item \textbf{S3:} Details of methodology for corpus lemmatisation and preparation.
\item \textbf{S4:} Details regarding data analysis.
\item \textbf{S5:} Details regarding phonological classes for irregulars.
\item \textbf{Figure S1:} Heap and Zipf distributions of verbs in CoHA prior to confining.
\item \textbf{Figure S2:} Verb tokens per decade prior to and after confining.
\item \textbf{Figure S3:} Frequency histogram of root types by category.
\item \textbf{Figure S4:} Temporal trajectory of six transitioning roots.
\item \textbf{Figure S5:} Percentage of past tense tokens per decade which are irregular.
\item \textbf{Figure S6:} Variance over time of items in irregular phonlogical classes.
\item \textbf{Table S1:} Summary of verb token counts per decade prior to and after corpus preparation.
\item \textbf{Table S2:} Summary of verb types by category per decade.
\item \textbf{Table S3:} Summary of root types by category per decade.
\item \textbf{Table S4:} Summary of phonological classes for irregular verbs.
\item \textbf{File S7:} File detailing verb types in each decade with their $I$ value and adjusted frequency (after confining; .csv format). Available at
\url{http://www.plosone.org/article/fetchSingleRepresentation.action?uri=info:doi/10.1371/journal.pone.0102882.s002} or upon request at  \href{mailto:ccuskley@gmail.com}{ccuskley@gmail.com}.

\item \textbf{File S8:} File detailing root types in each decade with their $I$ value and adjusted frequency (after confining; .csv format). Available at
\url{http://www.plosone.org/article/fetchSingleRepresentation.action?uri=info:doi/10.1371/journal.pone.0102882.s003} or upon request at  \href{mailto:ccuskley@gmail.com}{ccuskley@gmail.com}.

\end{itemize}

\clearpage









\section*{S2: Glossary}
\begin{itemize}
  \item {\bf Token}: A single instance of a verb use in the corpus, e.g., in ``I walked, she walks and he talks'' there are three verb tokens (\textit{walked, walks, talks}).

\item {\bf Type}: All token instances of a given verb lemma, e.g., in ``I walked, she walks and he talks'', there are two types, \textit{walk} and \textit{talk}. \textit{Walk} has two tokens (\textit{walked, walks}), while talk has one (\textit{talks}). The number of tokens for a given type determines its frequency.

\item {\bf Verb Type}: All separate verb lemmas, regardless of shared basic root verb. For example, \textit{do} and \textit{undo} constitute two separate verb types with different token counts.

\item {\bf Root Type}: Lemmas collapsed by basic root verb. When referring to root types, tokens of \textit{do} and \textit{undo} both contribute to the frequency of the root type \textit{do}.

\item {\bf $f$}: The basic usage frequency of a type, calculated as the number of tokens (from all tenses) divided by the total number of tokens in a decade (this number is held constant across decades at the value of 2,177,456 after confining at the size of the first decade, see text).

\item {\bf $f_{past}$}: The frequency of usage for a lemma in the past tense, defined as the total number of past tense tokens for the lemma divided by the size of the decade (overall number of tokens in decade) prior to confining (shown in Table S\ref{tab:removal}).

\item {\bf $I$}: The proportion of irregularity for a given type or class, defined as the number of irregular (non \textit{-ed}) past tense tokens for the type divided by the total number of past tense tokens for the type. $I$ can be defined only for types which exhibit a sufficient frequency (see \textit{Undefined}). Each type may have a different $I$ in each decade.

\item {\bf Undefined}: A verb or root is considered undefined in a given decade if its $f_{past} < 2.75 \cdot 10^{-6}$. In other words, a type's regularity is undefined if it has extremely low $f_{past}$.

\item {\bf Extended Vocabulary}: Entire set of verb or root types present after confining the corpus size to control for the increase in available text over time. This set includes verbs which enter and/or exit the vocabulary at any point during the 160 year time period.

\item {\bf Core Vocabulary}: The set of verb and root types present in all sixteen decades considered. Note that verbs or roots with very low $f$, though part of the core vocabulary, may have undefined $I$ in some decades.

\item {\bf Mostly Regular}: A type is classified as mostly regular if its $I < 0.5$.

\item {\bf Mostly Irregular}: A type is classified as mostly irregular if its $I \geq 0.5$.

\item {\bf $\epsilon$}: The error threshold for considering a verb or root to be regular or irregular (see below), set at $\epsilon = 0.01$. In other words, if the $I$ of a type is within $\epsilon$ of $0$ or $1$, it is considered regular or irregular (respectively). Note that a verb which is regular (irregular) in a decade may not stay regular (irregular) in other decades.

\item {\bf Stable Regular}: A verb or root is considered a stable regular if it respects $I \geq 0+\epsilon$ in all 16 decades.

\item {\bf Stable Irregular}: A verb or root is considered a stable irregular if it respects $I \geq 1-\epsilon$ in all 16 decades.

\item {\bf Active}: A verb or root is considered active if its $I$ exhibits a value respecting $0 + \epsilon \leq I \leq1-\epsilon$ in at least one decade.

\end{itemize}
\clearpage
\section*{S3: Corpus Preparation}
This section provides basic information about the corpus, and details regarding our methods in the preliminary stages of preparing the corpus for analysis. The Corpus of Historical American English contains over 400 million words of written English from 1810-2009. Each decade is genre balanced to contain roughly equal representation of fiction and non-fiction sources \cite{davies2012}. CoHA provides tagged frequency lists of words for each decade. 

Although CoHA spans 1810-2009, we used only the period between 1830 and 1989. The final two decades were removed (1990-2009) due to the fact that they are duplicated in the larger Corpus of Contemporary American English (CoCA; used for a separate investigation not reported here). The first two decades were removed as they displayed rather extreme growth in the database size. The number of verb tokens in the first decade (1810-1819) is approximately 20\% of the size of the second (1820-1829), and the second decade is still only about half the size of the third (1830-1839). Thereafter, growth levels off, with more moderate increases in the number of tokens between decades (see Figure S3, Table S2). We thus discarded the first two decades due to their small relative size; further potential effects of increasing database size were addressed by removing extremely low frequency items and by confining the size of each decade according to the 1830-1839 decade. This is explained in greater detail below.

\subsection*{Lemmatisation, Removals \& Confining}
Prior to analysis, the corpus was confined only to verbs and hand lemmatised by the authors (i.e., the words \textit{walked}, \textit{walking} and \textit{walks} were all tagged as tokens of the lemma \textit{walk}). In the process of lemmatisation, several types of removals were made. First, all modal auxiliary verbs in all tenses (e.g., \textit{can, could, may, must}) were removed as they are considered function words rather than lexical verbs (and are excluded from earlier studies of regularity, e.g., \cite{lieb2007}). Tagging errors, spelling and OCR errors, and items which occurred at extremely low frequency in very few decades were also removed. 

Hand lemmatisation allowed the coders to check many obvious tagging errors against their context in the corpus and remove such errors altogether. For example, \textit{chung} is tagged as a past tense verb, but invariably occurs as a proper noun (e.g., in a context such as ``I told her very briefly Chung Bong's story'' \cite{davies2012}). Note that not all tagging errors were removed; only those which were unknown English words and which had low enough frequency to allow checking against the corpus itself to verify the error\footnote{Although it is worth noting that tagging errors are more likely for low frequency items amenable to verification \cite{davies2012}.}. Where a past tense verb had the potential for error (e.g., \textit{abandoned} can be either an adjective or verb, and adjective tokens were sometimes incorrectly tagged as verbs as in ``this dear abandoned innocent'' \cite{davies2012}), but this error was not uniform (i.e., most tokens of \textit{abandoned} tagged as the past tense were correctly tagged), all tokens were included. This is due to the fact that manually checking all contexts of a particular word form is infeasible in CoHA (due to copyright restrictions at the time of analysis) for words which do not have a low token count.

Spelling or optical character recognition (OCR) errors were corrected where possible, for example \textit{abandonedthe} was coded as a past tense token of \textit{abandon}, and \textit{aaserting} is a clear instance of a misspelling of \textit{asserting}. Where these types of errors could not be straightforwardly interpreted, (e.g., for stranded bound morphems like \textit{-ify}, where each occurence was an error of the end of a different word, such as \textit{qual-ify, sign-ify}), they were removed entirely. Verbs were also removed if they did not occur with a frequency of more than $10^{-8}$ in at least three decades. In other words, as our analysis aimed to observe changes over time, we discarded verbs with both extremely low frequency \textit{and} a very short lifespan (\textless 30 years total, though not necessarily consecutive) in the corpus. 

These three criteria for removal led to the removal of between $9-12\%$ of the verb tokens in each decade (see Table S2). Figure S3 contrasts the token count of all verbs in the corpus (Davies, personal communication) versus the token count after our removals from each decade. The removal of modals accounted for the loss of between $6-8\%$ of tokens (between $55-80\%$ of all removals), depending on the decade. This percentage drops over time, indicating that the proportion of modals as a proportion of all verbs drops over time. This is likely due to the growth in the number of new lexical verbs evident even with the corpus size constrained (see main text, Figure 1). Additionally, the percentage of removal due to other criteria also increases over time (see Table S2); this is due to an increase in tokens removed due extremely low frequency items which occur for very short periods of time. This is consistent with an increase in the database size (more tokens can lead to the introduction of more low frequency types \cite{gerlachaltmann2013}), rather than any major variations in tagging errors, estimated to be stable around $1-2\%$ overall \cite{davies2012}. Moreover, the direction of these removals is conservative with respect to our results; in other words, even though a larger proportion of verb tokens were removed from the later decades, we still observed a growth in the number of verbs over time (see main text).

An increase in database size has the potential to affect the number of types observed \cite{gerlachaltmann2013}. Despite having removed very briefly or sporadically occurring low frequency items in the coding process, the growth in the number of tokens over time is still considerable (see Figure S3). To consider genuine vocabulary growth (rather than the growth in text available for digitized corpora over time), we confined the number of verbs considered in each decade to the number of tokens in the smallest decade (1830-1839; 2,177,456 tokens). This involved recreating a random sequence of all observed verb tokens in each decade. We then drew 2,177,456 tokens from each randomly sequenced set of verbs (with the exception of the first decade considered, 1830-1839, which remained intact). Verb types which remained after the set was confined constitute the extended vocabulary considered in our final analysis. Frequencies ($f$) were calculated based on the number of tokens per lemma divided by the size of the confined set. 
\clearpage
\section*{S4: Data Preparation}
After confining the size of each decade, we analysed several fundamental properties of the data: entrances and exits, frequency, root types, and regularity. Entrances and exits were considered in terms of single decade interval, meaning \textit{every} entrance and exit was counted, even if the same verb entered and then exited in consecutive decades. In other words, the verbs entering between 1840 and 1850 are verbs appearing in 1850 which did not appear in 1840, regardless of whether they appeared in 1830. However, overall, more verbs enter and stay (or enter more times than they exited), making the net result a growth in the number of verbs (see Figure 1E, main text).

\subsection*{Regularity}
Types in the extended vocabulary were categorized according to their proportion of irregularity ($I$), \textit{i.e.}, the fraction of irregular simple past tense occurrences over the total number of past tense tokens. The $I$ was calculated from our lemmatized version of the verb set, prior to confining decade size. Since the purpose of confining was to control for frequency effects related to database size, the process of confining did not recreate a \textit{tagged} database of verb tokens; information regarding past tense regularity is only available from the original lemmatized version. The $I$ was calculated using only the simple past tense; irregularity in the past participle was not considered, such that e.g., \textit{prove} is entirely regular (e.g., \textit{I proved her wrong}) although it has an irregular past participle (e.g., \textit{It has proven difficult}) in common usage. Irregular spellings such as \textit{paid} for the past tense of \textit{pay} (as opposed to \textit{payed}, which also occurs in the corpus) were considered regular past tense tokens, since spelling irregularity and variation were not considered in our analysis. Figure S5 shows the proportion of irregular tokens overall, which indicates that between $65-70\%$ of all past tense utterances are irregular. Even with the removal of the highest frequency irregulars, \textit{be} and \textit{have} (which also have the potential to be function verbs), around $50\%$ of all past tense verb tokens are irregular. This indicates that while regularity dominates types, irregularity dominates tokens.

We considered regularity undefined if past tense usage was so infrequent (or non-existent) that the regularity of a verb could not be determined without the potential for error. Therefore, in order to have defined regularity, a verb had to have past tense usage greater than or equivalent to a frequency of $2.75 \cdot 10^{-6}$ in a given decade. Because past tense usage and irregularity is based on the unconfined corpus, past tense usage frequency was calculated according to the original lemmatized corpus size for each decade. This frequency threshold is equivalent to at least $6$ past tense tokens for the first decade (1830-1839), but the number of past tense tokens required to reach this threshold scales with the increase in corpus size (such that e.g., at least $14$ past tense tokens are required to pass the threshold in the final decade). Frequency of usage in the past tense scales with overall frequency, such that low frequency items are much more likely to be undefined.

For broad contrasts in the extended vocabulary, all verbs were classified as mostly regular or mostly irregular ($I < 0.5$ and $I \geq 0.5$, respectively). However, the remainder of the analysis leveraged the availability of a scalar $I$ by contrasting regular and irregular roots with active roots in the core vocabulary. Root types with an $I \leq 0+ \epsilon$ were labelled as regulars, root types with  $I \geq1- \epsilon$ irregulars. When considering decades separately, active types are only considered active in decades where their $I \geq 0+ \epsilon$ or $I \leq 1- \epsilon$, but are considered (ir)regular elsewhere. However, across the entire time period, stable regulars or irregulars are types with an $I \leq 0+ \epsilon$ or $I \geq 1- \epsilon$ in every decade. Consequently, types with a $I \geq 0+ \epsilon$ or $I \leq 1- \epsilon$ in at least one decade were labeled as active across the time period.

The extended vocabulary presented with 6885 unique verb types. In order to examine the contribution of genuinely new verbs, exclusive of the contribution of new verbs which used an existing verb root productively, verbs were collapsed by their roots. This was particularly important since the use of existing irregular verb roots contributed in part to the introduction of ``new'' irregular types in the period. To this end, each of the 6885 verb types was assigned a root. The vast majority of verbs were monomorphemic and thus identical to their roots; e.g., the verb \textit{usher} is identical to the root \textit{usher} in all decades. Even many multimorphemic verbs were also identical to their roots, since words were only classed by free \textit{verb} roots, as irregularity can only be ``inherited'' from a verb root \cite{Pinker_Prince_1994}. For example, the verbs \textit{slave} and \textit{enslave} constitute separate verbs as well as separate roots, since they derive from the noun \textit{slave}, and thus do not share a verb root. Collapsing verbs by their roots resulted in a reduction in the number of unique types in the extended vocabulary, to 5791. Tables S3 and S4 summarize types by decade in terms of verbs and roots, respectively. The values of $f$ and $I$ for roots were re-calculated with the number of root tokens over decade size (for $f$) and the total number of irregular past tense tokens over the total number of past tense tokens for the entire root ($I$)\footnote{This makes it possible that in the case of the number of active types in a single decade, there are more active roots than active verbs. This is because using multiple verbs to create a single $I$ drives some additional roots into transition.}.

Of 200 unique irregular verb types (151 root types) in the corpus, 18 (9\%) of these appear after 1840, while 22 (11\%) are lost, making for a small net decrease in the number of irregular verb types (the remaining 80\% are in the core vocabulary). In the case of regulars, the birth rate observed is not only much greater than that of irregulars, but it dwarfs the death rate; 26.3\% of all regular verb types in the corpus are born after 1840, while only 14.3\% of verbs are lost. 

Calculation of root types shows that the 18 entrances of irregular verbs occur either because of definition or root proliferation (i.e., the productivity of an existing irregular root, as in \textit{do-undo}). Definition occurs when a verb acquires sufficient frequency of usage in the past tense for its regularity to be reliably defined; i.e., it moves out of the undefined category. Four of the 18 entering irregulars (approximately 23\%) are undefined in the early decades, but enter as mostly irregular ($I \geq 0.5$) at their first occurrence and remain mostly irregular throughout their lifetime. The largest percentage of irregular verb birth is accounted for by the proliferation of irregular roots; 11 of the 18 nascent irregulars  (just over 60\%) are multi-morphemic verbs using an existing irregular root, such as \textit{outdo} and \textit{override}. The remaining three new irregular verbs (constituting 17\%), are not entrances, rather, they are instances of irregularization: verbs which at their first occurrence were regular, but by the final decade have become irregular. If verbs are collapsed by their roots - eliminating the process of root proliferation as a mechanism of birth, this leaves only 7 new irregular roots (of 151 unique root types total): 3 irregularizations and 4 definitions.

Unlike for irregulars, root proliferation is not a major contributing force behind new regular types, since collapsing regular types into roots has little effect on the observed rates of birth and death (adjusting them to 26\% and 13.8\%, respectively). Definition accounts for a large proportion of new regular verb types, with 78.7\% of new regulars becoming defined as regular sometime after 1840 (although they occur with some $f$ in early decades). Verbs entering the system form the second largest source of new regulars, accounting for almost 21\% of new regulars. Lastly, regularization accounts for a small minority of new regulars, with only three verbs regularizing completely, constituting less than  0.5\% of regular verb growth. In other words, defining and entering verbs skew drastically towards being regular, and a growth of the number of types over time is the primary force driving an overall increase in regularity in the language system.
\clearpage
\section*{S5: Phonological classes}
All verbs which had a non-zero $I$ at any time during the 160 years examined were classed according to the change from the infinitive form to the irregular past tense form. A full list of all 52 classes and their members is provided in Table S5. Verbs which exhibited multiple irregular forms were phonologically classified based on their most frequent irregular form (i.e., \textit{swing} was classed with \textit{ring} instead of \textit{string}, although the form \textit{swung} did occur in a minority). Suppletive forms such as \textit{go} and \textit{be} were in their own class, and forms such as \textit{slay/slew} and \textit{lay/lay} did not class identically with any other verbs, and were thus also classed alone. Each class has an $f$ defined as the sum of the frequencies of its root members, while the $I$ in each class is calculated by dividing the sum of irregular past tense tokens in the class by the sum of all past tense tokens.

These classes may not be optimal, and could in fact be more or less fine grained. For example, this metric is coarse in that it does not take into account the presence of several complex onsets in irregulars (e.g., \textit{spr-, str-}), which may effect irregularisation (in other words, it does not use overall phonological distance; see \cite{albrighthayes2003}). On the other hand, most systems of classification are much broader with as few as 6-8 classes encompassing all irregular verbs \cite{greenbaumquirk1996,mclelpatterson2002}.

Because an irregular form is required for class membership, class sizes are not fixed over time. In other words, while irregulars have the potential to contribute to classes, regulars do not (such that, e.g., there is no subset of regular classes with phonologically similar members). When a particular root type regularizes completely, it leaves its irregular class entirely; in other words, the class can be said to be losing a member to regularization. For example, the verb \textit{work} occurs early on in the corpus with an $I$ of $0.04$ as there is still some usage of the irregular form \textit{wrought}. While \textit{work} has this positive $I$, it belongs to the \textit{teach} class. However, by 1930, \textit{work} has an $I$ of $0$, and has therefore left the \textit{teach} class. Thus, in 1930, the \textit{teach} class shrinks in overall size, and \textit{work} no longer contributes to either the class's $f$ or its $I$.  Likewise, each class also has the potential to gain members as new irregular forms emerge (or new verbs enter as irregular, although this is rare in our data). For example, the verb \textit{ruin} occurs only with the form \textit{ruined} until 1880, at which point the form \textit{ruint} emerges. As \textit{ruin} now has an $I$ of $0.02$, it enters to the \textit{burn} class contributing to both its $f$ and $I$.
\clearpage

\noindent \textbf{Figure S1: Heaps curves (A) and Zipfian distributions (B) prior to confining the dataset to the size of the 1830-1839 decade.}
\begin{center}
  \includegraphics[width=.5\textwidth]{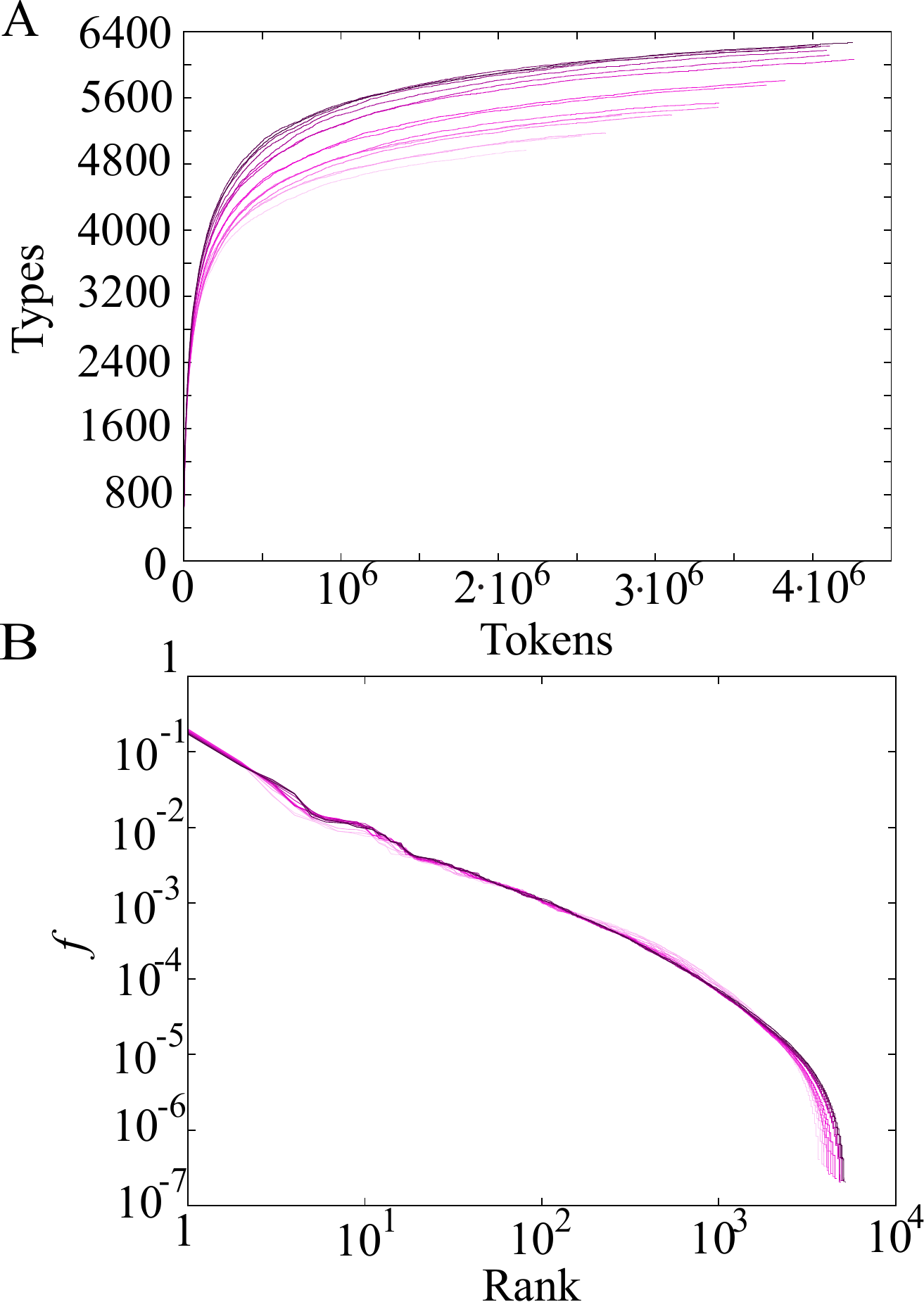}
\end{center}
\clearpage
\noindent {\bf Figure S2. Number of verb tokens in CoHA in each decade.} The red line indicates the number of all tokens prior to removal of modal auxiliaries, obvious errors, and extremely low frequency/sporadically occurring types, while the blue line indicates the number of tokens after this removal.
\begin{center}
  \includegraphics[width=.7\textwidth]{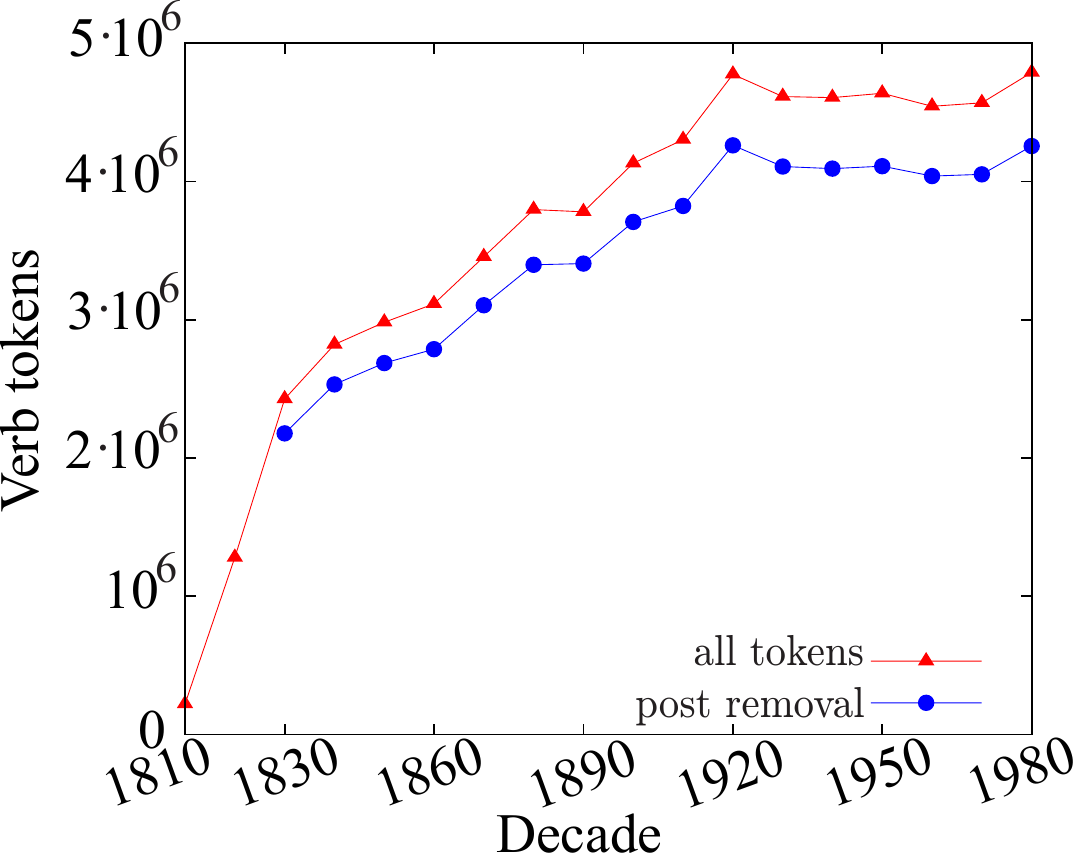}
\end{center}
\clearpage
\noindent \textbf{Figure S3. Depicts the frequency histogram of root types divided by category} (regular, irregular, undefined and active) in four different decades (compare to Figure 1E in the main text, which depicts mostly regular and mostly irregular root types). This shows that the growth in number of types is mainly a consequence of entering regular types, many of which were simply previously undefined. The starting point of the first frequency bin for each decade is indicated.
\begin{center}
 \includegraphics[width=.8\textwidth]{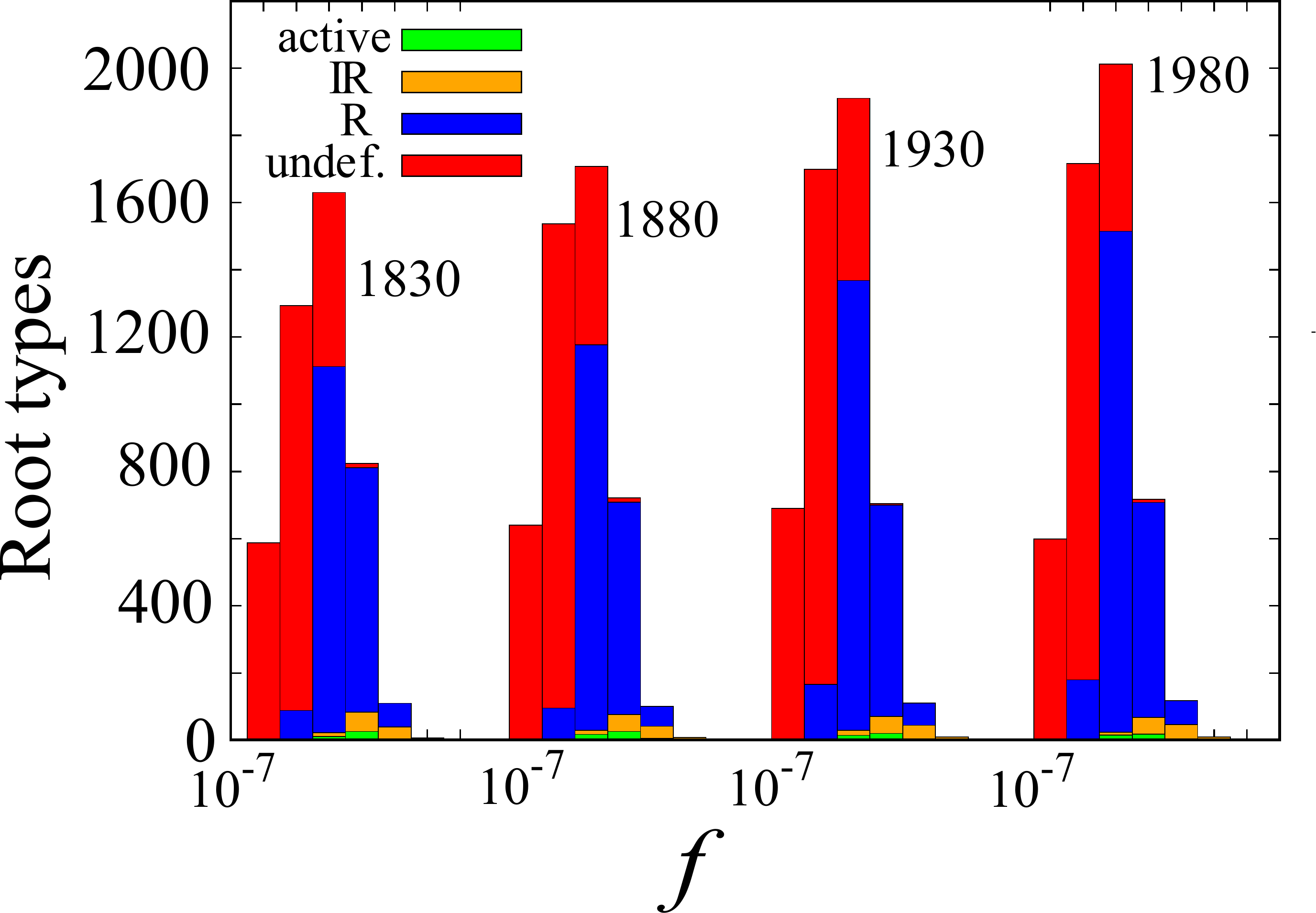}
\end{center}
\clearpage
\noindent \textbf{Figure S4. Transitioning roots.} The six roots in the database which transition from (A) mostly irregular to mostly regular, and (B) mostly regular to mostly irregular.
\begin{center}
  \includegraphics[width=\textwidth]{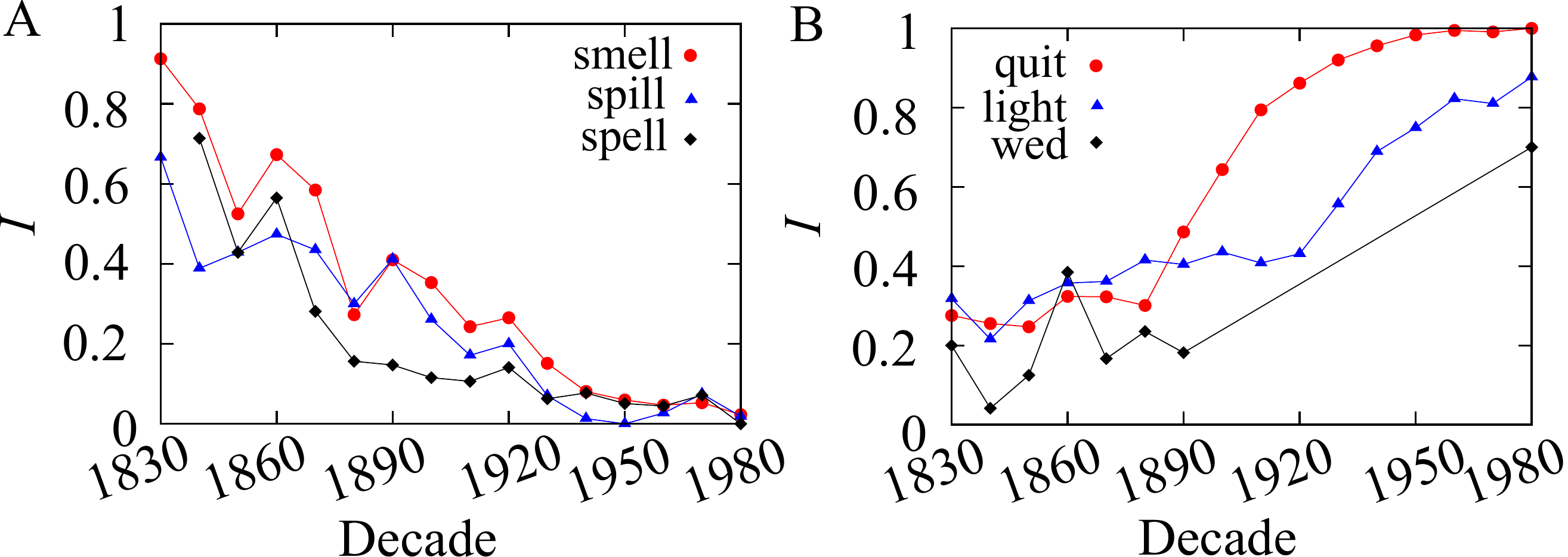}

\end{center}
\clearpage
\noindent \textbf{Figure S5. Proportion of irregular past tense tokens}, $I_{tot}$, in each decade for all verbs (red line) and excluding excluding particularly high frequency types ``be'' and ``have'' (blue line).
\begin{center}
  \includegraphics[width=.7\textwidth]{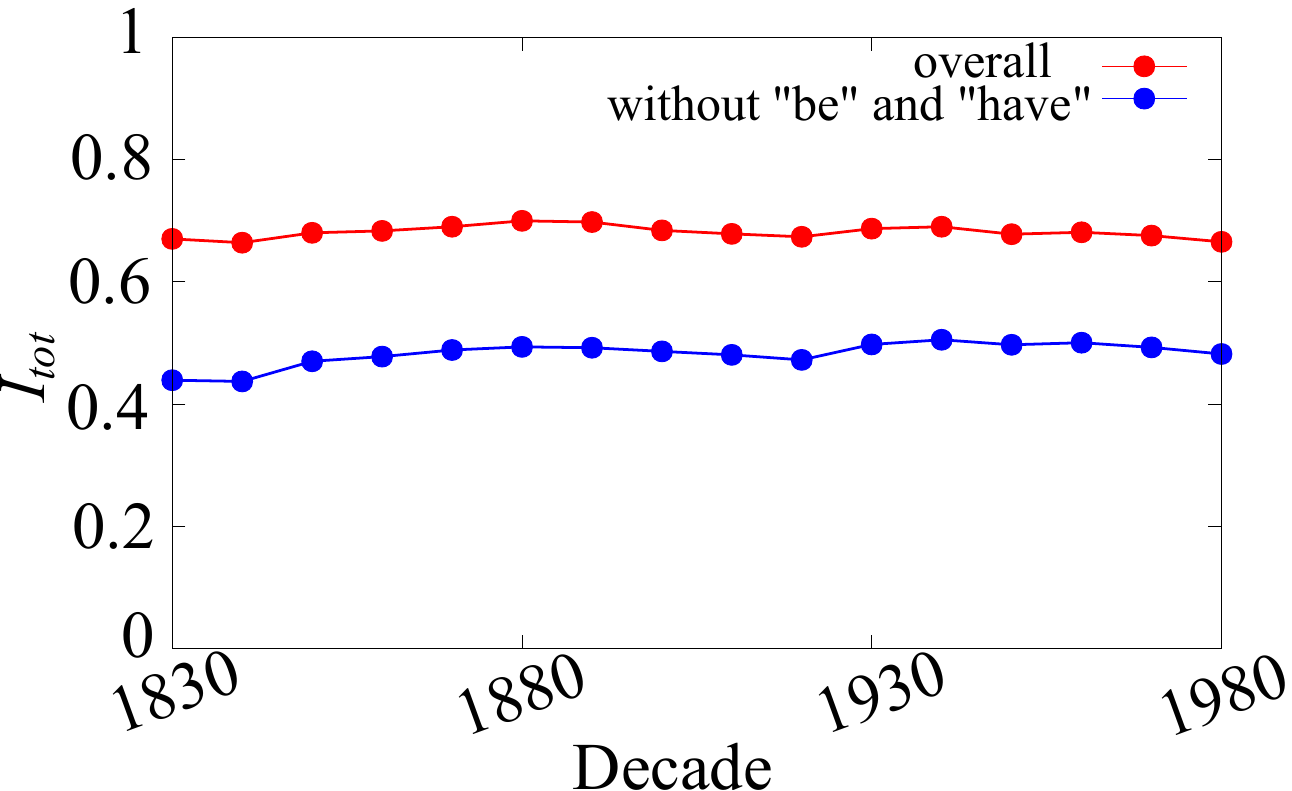}
\end{center}
\clearpage
\noindent {\bf Figure S6. Plot of $I$ versus the $f_{sum}$ for each phonological class over time.} Time is represented by color (red hue for the first decade, blue hue for the last decade). Each member of a class is plotted individually (with its own $I$ and $f$) and connected to the class by a line. The size of the circle depicting the class indicates how many members are in the class.
\begin{center}
  \includegraphics[width=\textwidth]{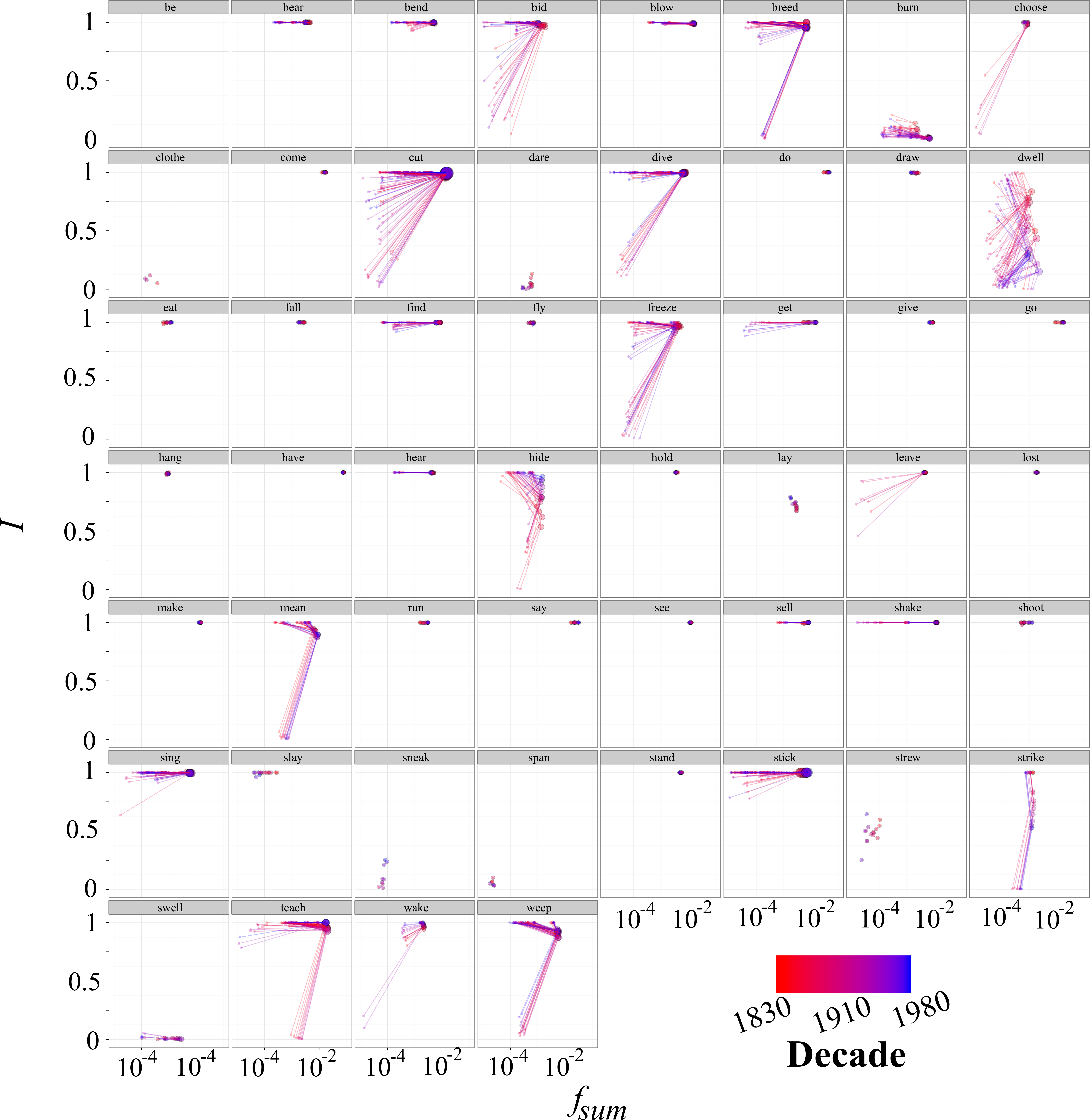}

\end{center}
\clearpage
\noindent \textbf{Figure S7. Visualization of the variance in regularity among classes.} In each decade, for a given class with a size, $s$, we can define its variance, $\sigma$, as:

$$
\sigma = \frac{ \sqrt{ \sum_{v} (I_r - I_c)^2 } }{\sqrt{s}} \ ,
$$
Where $I_r$ is the proportion of irregularity of each member root of the class, and $I_c$ is the overall $I$ of the class. This figure shows a plot of the variance of $\sigma$ over time against the average $\sigma$ ($\bar \sigma$) for each class. Classes with higher $\bar \sigma$ are in the process of losing a member throughout the time period, while higher variance in $\sigma$ indicates the loss of a member. For example, the \textit{mean} class is slowly losing \textit{dream} and \textit{lean}, which have an $I$ of $0.108$ and $0.007$ respectively by the final decade. The \textit{choose} and \textit{strike} classes both lose members (\textit{behoove} and \textit{climb}, respectively) in the time period. Classes with low variance in $\sigma$ over time and low $\bar \sigma$ are highly stable (and/or have only a single member, e.g., the \textit{be} class).The variance of $\sigma$ over time versus $\bar \sigma$ for
  all classes. Only classes with a variance over time in $\sigma  \textgreater  0.05$ are labelled.

\begin{center}
  \includegraphics[width=.5\textwidth]{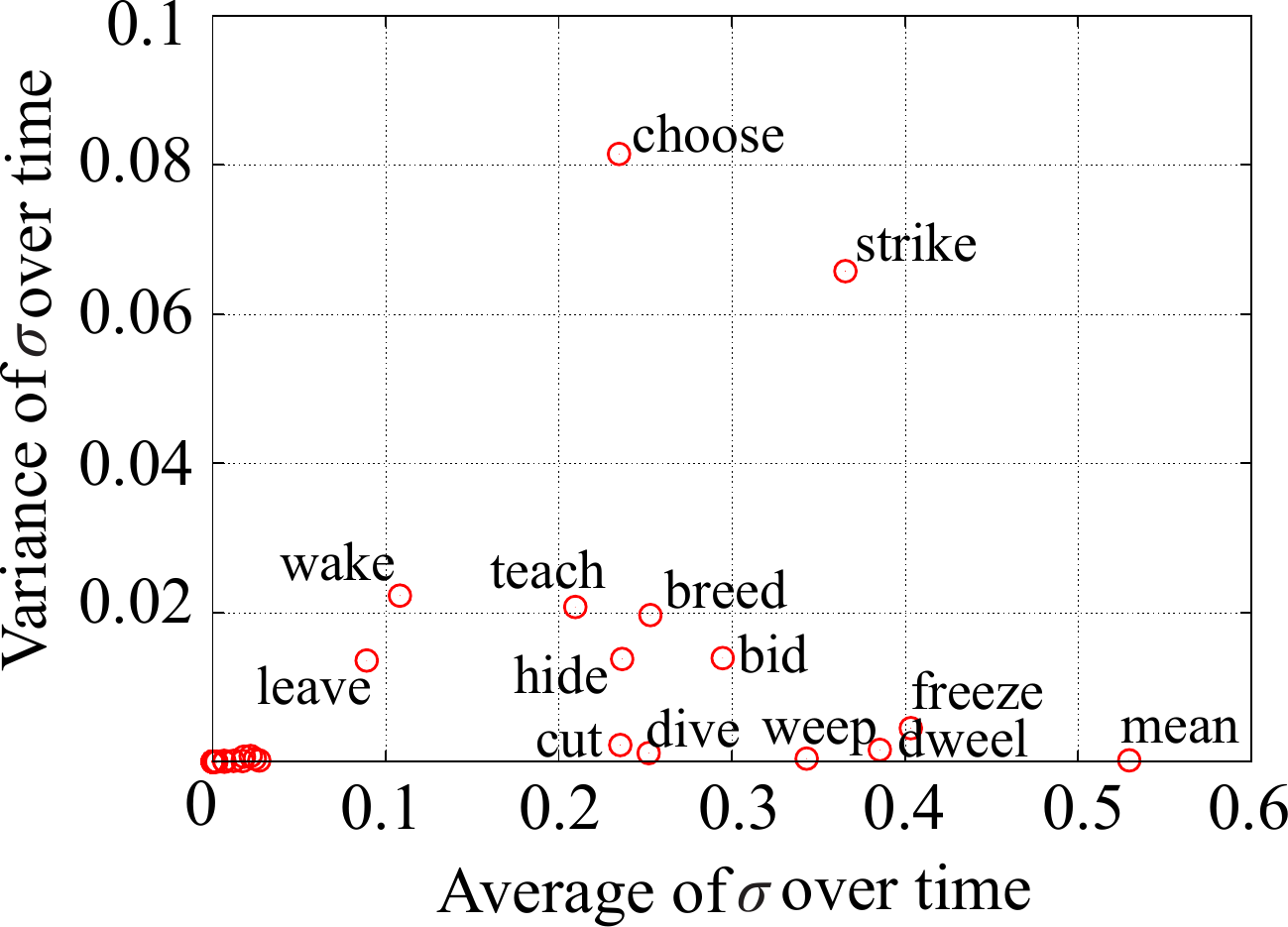}
\end{center}
\clearpage
\begin{table*}[h!]
\begin{center}
\caption{{\bf Summary of verb tokens removed from the original, unconfined CoHA set.} Percentages show removals as a percentage the original number of tokens.}\label{tab:removal}
\begin{tabular}{|cx{2.2cm} cx{2.2cm} cx {2.2cm}cx{2.2cm}cx{2.2cm}cx{2.2cm}}
\hline
Decade &	All Verb Tokens &	Post Removals & Removed & Modals Removed &  Other Removals\\ \hline
    1830-1839 &	2426234 &	2177456 &	10.25\% & 8.34\%&	1.91\%\\
    1840-1849 &	2820505 &	2531984 &	10.23\%& 8.29\% &	1.94\%\\
    1850-1859 &	2981280 &	2686698 &	9.88\% & 8.35\% &	1.54\%\\
    1860-1869 &	3114518 &	2787135 &	10.51\% & 8.14\% &	2.38\%\\
    1870-1879 &	3454665 &	3104851 &	10.13\%& 8.11\% &	2.02\%\\
    1880-1889 &	3795539 &	3397440 &	10.49\% & 8.11\% &	2.38\%\\
    1890-1899 &	3780288 &	3405447 &	9.92\% & 7.65\% &	2.27\%\\
    1900-1909 &	4131287 &	3706878 &	10.27\% & 7.49\% &	2.78\%\\
    1910-1919 &	4301527 &	3823067 &	11.12\% & 7.38\% &	3.74\%\\
    1920-1929 &	4773647 &	4260603 &	10.75\%& 7.04\% &	3.71\%\\
    1930-1939 &	4612946 &	4107362 &	10.96\% & 6.93\% &	4.03\%\\
    1940-1949 &	4605199 &	4092289 &	11.14\%& 6.86\% &	4.28\%\\
    1950-1959 &	4636677 &	4110181 &	11.36\% & 6.83\% &	4.53\%\\
    1960-1969 &	4543216 &	4037815 &	11.12\% & 6.77\% &	4.36\%\\
    1970-1979 &	4567916 &	4050662 &	11.32\% & 6.61\% &	4.72\%\\
    1980-1989 &	4787625 &	4255111 &	11.12\% & 6.33\% & 4.79\%\\
    \hline
\end{tabular}
\end{center}
\end{table*}

\clearpage
\begin{table*}[htdp]
\caption{\textbf{Summary of verb type categories across decades}}\label{tab:verb_types}
\begin{center}
\begin{tabular}{|cx{50pt}cx{25pt}cx{1.5cm}cx{1.5cm}cx{25pt}cx{1.5cm}cx{1.5cm}cx{1.5cm}cx{1.5cm}}
\hline
Decade &	Mostly Irregular &  Mostly Regular & Irregular & Regular & Active & Entering In & Exiting After& Undefined \\ \hline
    1830-1839 &	172 & 2061 & 148 &	2045 &	40 & --- &	295 &	2731 \\
    1840-1849 &	172 & 2149 & 149 &	2130 &	42  & 389 &	335 &	2737\\
    1850-1859 &	171 & 2080 & 152 &	2064 &	35  & 355 &	256 &	2827\\
    1860-1869 & 174 & 2181 & 147 &	2161 &	47  & 473 &	382 &	2940\\
    1870-1879 &	172 & 2126 & 150 &	2107 &	41  & 319 &	337 &	2934\\
    1880-1889 &	171 & 2038 & 144 &	2012 &	53  & 373 &	346 &	3059\\
    1890-1899 &	175 & 2096 & 153 &	2072 &	46  & 400 &	319 &	3051\\
    1900-1909 &	173 & 2144 & 150 &	2122 &	45  & 504 &	357 &	3190\\
    1910-1919 &	170 & 2189 & 147 &	2168 &	44  & 395 &	267 &	3186\\
    1920-1929 &	171 & 2332 & 147 &	2311 &	45  & 499 &	355 &	3274\\
    1930-1939 &	172 & 2290 & 150 &	2271 &	41  & 391 &	336 &	3351\\
    1940-1949 &	169 & 2327 & 146 &	2306 &	44  & 380 &	359 &	3361\\
    1950-1959 &	168 & 2347 & 144 &	2332 &	39  & 407 &	356 &	3390\\
    1960-1969 &	169 & 2314 & 150 &	2296 &	37  & 399 &	355 &	3465\\
    1970-1979 &	171 & 2334 & 149 &	2315 &	41  & 345 &	336 &	3433\\
    1980-1989 &	168 & 2475 & 146 &	2461 &	36  & 356 & --- &	3315\\
    \hline
\end{tabular}
\end{center}
\label{default}
\end{table*}
\clearpage
\begin{table*}[htdp]
\caption{\textbf{Summary of root type categories across decades}}\label{tab:root_types}
\begin{tabular}{|cx{50pt}cx{25pt}cx{1.5cm}cx{1.5cm}cx{25pt}cx{1.5cm}cx{1.5cm}cx{1.5cm}cx{1.5cm}}
\hline
Decade &	 Mostly Irregular &  Mostly Regular  &  Irregular  &	 Regular  & Active & Entering In & Exiting After&Undefined\\ \hline
    1830-1839 &	138 & 1992 & 112 &	1976 &	42 & --- &	243 & 2323\\
    1840-1849 &	138 & 2071 & 112 &	2053 &	44 & 308 &	274 & 2309\\
    1850-1859 &	136 & 2012 & 117 &	1996 &	35 & 281 &	187 & 2377\\
    1860-1869 & 137 & 2109 & 112 &	2088 &	46 & 391 &	303 & 2483\\
    1870-1879 &	135 & 2048 & 115 &	2027 &	41 & 240 &	252 & 2483\\
    1880-1889 &	134 & 1959 & 109 &	1933 &	51 & 303 &	287 & 2624\\
    1890-1899 &	134 & 2026 & 113 &	2002 &	45 & 297 &	235 & 2567\\
    1900-1909 &	137 & 2078 & 116 &	2056 &	43 & 405 &	272 & 2682\\
    1910-1919 &	135 & 2126 & 113 &	2104 &	44 & 284 &	175 & 2648\\
    1920-1929 &	136 & 2257 & 113 &	2235 &	45 & 366 &	250 & 2707\\
    1930-1939 &	137 & 2219 & 113 &	2199 &	42 & 275 &	244 & 2769\\
    1940-1949 &	135 & 2253 & 112 &	2232 &	44 & 250 &	246 & 2743\\
    1950-1959 &	136 & 2279 & 113 &	2263 &	39 & 284 &	249 & 2754\\
    1960-1969 &	136 & 2242 & 117 &	2223 &	38 & 270 &	246 & 2812\\
    1970-1979 &	137 & 2267 & 116 &	2247 &	41 & 233 &	242 & 2773\\
    1980-1989 &	135 & 2397 & 113 &	2382 &	37 & 238 &	--- & 2641\\
    \hline
\end{tabular}
\end{table*}
\clearpage

\begin{table}[h!]
\caption{\textbf{Summary of the phonological classes implemented}}\label{tab:classes}
\begin{tabular}{|m{3cm}|m{3cm}|}
\hline
Root Members 	&Change\\	\hline
\textbf{be} & be $\rightarrow$ went (suppletive) \\	\hline
\textbf{bear}, swear, tear, wear	& \textipa{/E/} $\rightarrow$ \textipa{/o/}\\	\hline
\textbf{bend}, build, lend, send, spend	& \textipa{/d/} $\rightarrow$ \textipa{/t/}\\ \hline
\textbf{bid}, braid, rid, shed, sled, spread, wed	 & \textipa{/d/} $\rightarrow$ \textipa{/d/} $+\emptyset$\\	\hline
\textbf{blow}, grow, know, throw &	\textipa{/oU/} $\rightarrow$ \textipa{/u/}\\	\hline
\textbf{breed}, bleed, feed, lead, meet, plead, read, speed&	\textipa{/i/} $\rightarrow$ \textipa{/E/} \\	\hline
\textbf{burn}, drown, learn, ruin  & \textipa{/n/} $\rightarrow$ \textipa{/nt/}\\	\hline
\textbf{choose}, behoove&	\textipa{/u/} $\rightarrow$ \textipa{/oU/}\\	\hline
\textbf{clothe} & \textipa{/oUD/} $\rightarrow$ \textipa{/\ae d/}\\	\hline
\textbf{come} & \textipa{/2/} $\rightarrow$ \textipa{/\ae/}\\	\hline
\textbf{cut}, beat, bet, burst, bust, cast, cost, hit, hurt, knit, let, put, quit, set, shut, slit, split, thrust, wet	&	\textipa{/t/} $\rightarrow$ \textipa{/t/} $+\emptyset$\\	\hline
\textbf{dare} & dare $\rightarrow$ durst (suppletive)\\	\hline
\textbf{dive}, drive, ride, rise, shine, smite, stride, strive, write & \textipa{/aI/} $\rightarrow$ \textipa{/oU/} \\	\hline
\textbf{do}&do $\rightarrow$ did (suppletive) \\	\hline
\textbf{draw}	&\textipa{/a@/} $\rightarrow$ \textipa{/u/} \\	\hline
\textbf{dwell}, heal, kneel, scare, smell, spill, spell, spoil	& alveolar approximant $+$ \textipa{t}\\	\hline
\textbf{eat} & \textipa{/i/} $\rightarrow$ \textipa{/\ae/}\\	\hline
\textbf{fall}	& \textipa{/a@/} $\rightarrow$ \textipa{/E/}\\	\hline
\textbf{find}, bind, grind, wind	&	\textipa{/aI/} $\rightarrow$ \textipa{/aU/}\\	\hline
\textbf{fly}	&\textipa{/aI/} $\rightarrow$ \textipa{/u/}\\	\hline
\textbf{freeze}, heave, speak, squeeze, steal, weave	&\textipa{/i/} $\rightarrow$ \textipa{/oU/}\\	\hline
\textbf{get}, tread	 & \textipa{/e/} $\rightarrow$ \textipa{/A/} \\	\hline
\textbf{give}&	\textipa{/I/} $\rightarrow$ \textipa{/eI/} \\	\hline
\end{tabular}
\begin{tabular}{|m{3cm}|m{3cm}|}
\hline
Root Members 	&Change\\	\hline
\textbf{go}	& go $\rightarrow$ went (suppletive)\\	\hline
\textbf{hang}  & \textipa{/\ae/} $\rightarrow$ \textipa{/2/}\\	\hline
\textbf{have}	&\textipa{/v/} $\rightarrow$ \textipa{/d/}\\	\hline
\textbf{hear}, flee& \textipa{/i/} $\rightarrow$ \textipa{/e/} + word final \textipa{/d/}\\	\hline
\textbf{hide}, bite, light, slide 	&	\textipa{/aI/} $\rightarrow$ \textipa{/I/}\\	\hline
\textbf{hold}	& \textipa{/o/}  $\rightarrow$ \textipa{/e/}\\	\hline
\textbf{lay}	 & \textipa{/EI/} $\rightarrow$ \textipa{/EI/}\\	\hline
\textbf{leave}, cleave & \textipa{/iv/} $\rightarrow$ \textipa{/Eft/}\\	\hline
\textbf{lose}& \textipa{/u/} $\rightarrow$ \textipa{/A/} + word final \textipa{/t/}\\	\hline
\textbf{make}&	\textipa{/k/} $\rightarrow$ \textipa{/d/}\\	\hline
\textbf{mean}, deal, dream, feel, lean & \textipa{/i/} $\rightarrow$ \textipa{/e/} + word final \textipa{/t/}\\	\hline
\textbf{run}	& \textipa{/2/} $\rightarrow$ \textipa{/\ae/}\\	\hline
\textbf{say}	& \textipa{/EI/} $\rightarrow$ \textipa{/ed/}\\	\hline
\textbf{see}	 &\textipa{/i/} $\rightarrow$ \textipa{/a@/}\ \\	\hline
\textbf{sell}, tell	& \textipa{/E/} $\rightarrow$ \textipa{/o/} + word final \textipa{/d/}\\	\hline
\textbf{shake}, forsake, take	& \textipa{/EI/} $\rightarrow$ \textipa{/\oe/}\\	\hline
\textbf{shoot} &\textipa{/u/} $\rightarrow$ \textipa{/A/}\ \\	\hline
\textbf{sing}, drink, ring, shrink, sink, sit, spit, spring, stink, swim 	& \textipa{/I/} $\rightarrow$ \textipa{/\ae/}\\	\hline
\textbf{slay}	&	\textipa{/eI/} $\rightarrow$ \textipa{/u/}\\	\hline
\textbf{sneak}& \textipa{/i/} $\rightarrow$ \textipa{/2/}\\	\hline
\textbf{span}	 & \textipa{/n/} $\rightarrow$ \textipa{/n/}\\	\hline
\textbf{stand} & \textipa{/\ae/} $\rightarrow$ \textipa{/U/}\\	\hline
\textbf{stick}, begin, cling, dig, fling, sling, slink, spin, sting, string, swing, win, wring	& \textipa{/I/} $\rightarrow$ \textipa{/2/}\\	\hline
\textbf{strew}	&	\textipa{/u/} $\rightarrow$ \textipa{/u/}\\	\hline
\textbf{strike}, climb&	\textipa{/aI} $\rightarrow$ \textipa{/2/}\\	\hline
\textbf{swell}, help, step & \textipa{/e/} $\rightarrow$ \textipa{/o/}\\	\hline
\textbf{teach}, beseech, bring, buy, catch, fight, seek, think, work& Replace final syllable rime with \textipa{/a@t/} (Ruckumlaut)\\	\hline
\textbf{wake}, break	 & \textipa{/EI/} $\rightarrow$ \textipa{/oU/}\\	\hline
\textbf{weep}, creep, keep, leap, sleep, sweep &\textipa{/i/} $\rightarrow$ \textipa{/E/} + word final \textipa{/t/}\\ \hline

\end{tabular}
\end{table}

\newpage


\end{document}